\documentclass[intlimits,twoside,a4paper]{article}

\usepackage{amsmath}
\DeclareMathOperator{\Tr}{Tr}
\DeclareMathOperator{\spec}{spec}
\usepackage{amssymb}
\usepackage{graphicx}
\usepackage[cp1251]{inputenc}

\usepackage[eqsecnum]{cmpj3}

\issue{2017}{20}{3}{33001}
\doinumber{10.5488/CMP.20.33001}

\title[From the arrow of time in Badiali's quantum approach to the dynamic
meaning of Riemann's hypothesis]{From the arrow of time in Badiali's quantum approach to the dynamic
meaning of Riemann's hypothesis}
\author[P. Riot, A. Le M\'ehaut\'e]{P. Riot\refaddr{label1}, A. Le M\'ehaut\'e\refaddr{label1,label2,label3} }
\addresses{
\addr{label1}
Franco-Quebecois Institute, 37 rue de Chaillot, 75016 Paris, France
\addr{label2}
Physics and Information Systems Departments, Kazan Federal University,\\
18--35 Kremlevskaia St., 480 000 Kazan, Tatarstan, Russia Federation
\addr{label3}
Materials Design, 18 rue Saisset, 92120 Montrouge, France
}
\date{Received April 19, 2017, in final form June 8, 2017}

\begin{document}

\maketitle

\begin{abstract}
The novelty of the Jean Pierre Badiali last scientific works stems
to a quantum approach based on both (i) a return to the notion of
trajectories (Feynman paths) and (ii) an irreversibility of the quantum
transitions. These iconoclastic choices find again the Hilbertian
and the von Neumann algebraic point of view by dealing statistics
over loops. This approach confers an external thermodynamic origin
to the notion of a quantum unit of time (Rovelli Connes' thermal time).
This notion, basis for quantization, appears herein as a mere criterion
of parting between the quantum regime and the thermodynamic regime.
The purpose of this note is to unfold the content of the last five
years of scientific exchanges aiming to link in a coherent scheme
the Jean Pierre's choices and works, and the works of the authors
of this note based on hyperbolic geodesics and the associated role of
Riemann zeta functions. While these
options do not unveil any contradictions,  nevertheless they give birth to an intrinsic arrow of time different
from the thermal time. The question of the physical meaning of Riemann
hypothesis as the basis of quantum mechanics, which was at the heart of
our last exchanges, is the backbone of this note.

\pacs 05.30.-d, 05.45.-a, 11.30.-j, 03.65.Vf
\end{abstract}

\section{From algebraic analysis of quantum mechanics to ``irreversible''
Feynman paths integral}

Despite the unstoppable success of the technosciences based on both
quantum mechanics, standard particle model and cosmological model,
at least two questions must be investigated among many issues that
the theories leave open \cite{Penrose,smolin}: (i) the question
of the ontological status of the time and (ii) the obsessive interrogation
concerning the existence or the absence of an intrinsic ``arrow of time''.
The origin of these questions comes from the equivocal equivalence
of the status of time in any types of mechanical formalisms. For example,
within Newtonian vision, the observable $f$ can be analysed algebraically
using action-integral through the Lagrangian $L$ while Poisson brackets
gives time differential representations $\rd f/\rd t=\{H,f\}$. According
to Noether theorem, the energy, referred to the Hamiltonian $H$,
is no other than the tag of a time-shift independence of physical
laws, namely a compact commutativity. The statistical knowledge of
the high dimensions system requires (i) the definition of a Liouville
measure $\mu_{\text L}$ based on the symplectic structure of the phase
space and (ii) the value of the configuration distribution $Z_{\text C}$,
therefore $\rd\mu\sim(1/Z_{\text C})\,\re^{-\beta H}$, with $\beta=1/k_{\text B}T$ related
to the inverse of the temperature. This point of view is discretized
in quantum mechanics (QM).

With regard to quantum perspectives, mechanical formalism introduces
(i) a thickening of the mechanical dot, (ii) the substitution of real
variables through the spectrum of operators and (iii) an emphasis on
the role of probability. According to von Neumann, the stable core
of the operator algebra required to fit the quantum data must be based
upon groupoids acting on observables. In the Heisenberg framework,
the observable $\hat{f}$ (for instance the paradigmatic example of
the set of the rays of materials emissions) is represented by self-adjoint operators in Hilbert space $l^{2}(\mathbb{R}^{3},\mathbb{C}$)
which values can be reduced within Born-matrix representation to a
set of eigenvectors $|\varphi_{n}\rangle$ chosen in the spectrum $\spec(\hat{f})$
of the groupoid. Energy distribution is given through the linear relations
$\hat{H}|\varphi_{n}\rangle=E|\varphi_{n}\rangle$, where  the Hamiltonian $\hat{H}$
represents the energy self-adjoint operator. The dynamics is implemented
by using the commutator: $[\hat{H},\hat{f}]$ which replaces the Poisson
bracket, namely $\rd\hat{f}(t)/\rd t=\frac{2\piup \ri}{h}[\hat{H},\hat{f}]$.
The capability of giving cyclic representations of von Neumann algebra
(extended to Weyl non-commutative algebra for standard model) leads
to expressing the dynamics via the eigenvectors Fourier components $|\psi_{n}(t)\rangle=|\varphi_{n}\rangle\exp(-\ri E_{n}t/\hbar)$.
This representation is unitarily equivalent to a wave mechanics usually
expressed through the Schr\"odinger equation, $\ri\hbar\frac{\rd}{\rd t}\psi(r,t)=[-\frac{\hbar^{2}}{2m}\nabla^{2}+V(r,t)]\psi(r,t)$.
The shift from non-linear finite to linear infinite system must be
based upon the statistics dealing with a $\Lambda$-extension of the
system, through a linear and positive forms $\hat{f}\in A$ $\Phi_{\Lambda}(A)=(1/Z_{\text C})\Tr\exp(-\beta H_{\Lambda_{A}})$.
Hence, the average value of the observable $\hat{f}\in A$ is
a trace of an exponential operator. Usually, the distribution of physical
data must be given by a measure of probability on $\spec(A)$. Thus,
we cannot deal with QM without dealing with Gaussian randomness imposed
by some external thermostat. At this step, a useful notion is the
notion of density matrix given by: $\rho N=\exp(-\beta H)$. Unfortunately,
$N$ the normalization constant suffers from all misgivings involved
in thermodynamics, by the ``shaky'' notion
of equilibrium.

\footnote{The extension of $l^{2}(\mathbb{R}^{3},\mathbb{C}$) toward $l^{2}(\mathbb{R}^{3},\mathbb{C}^{2\times2}$)
shifts the second order equation onto a first order equation with
observables then based upon matrix values. This shift gives birth
to the Dirac operator whose algebra founds the spineur standard model
of physics. It is clearly based on inner automorphisms and internal
symmetries \cite{Connes-1,Connes 2}.}Each item of the above visions imposes its own algebraic constraints
but enforces a paradigmatic concept of time parameter \cite{Connes-1}
as a reversible ingredient of the physics. At this step, the statistics
appears as the only loophole capable of introducing irreversibility as
a path to an assumptive equilibrium state for finite $\beta$ value.
Nevertheless, as shown above, this assumption requires the $\Lambda$-extension,
namely, the transfer of the operator algebra in the framework of \textit{C{*}}-algebra
in which the $A$-algebra of its Hermitian elements patterns the
transfer (rays) between a set of perfectly well defined states.
Starting from the notion of groupoid and from the algebra of magma
upon the states and by analysing the symmetries, a mathematician
can also consider the equilibrium from a set of cyclic states $\Omega$
of $\hat{f}$, based on Gelfand, Naimark, Segal construction (\textit{GNS
construction)} \cite{GNS} binding quantum states and the cyclic states
(cyclic transfer which assumes a specific role of scalar operators,
called $M$-factors). At this stage, two points of view must be matched
together to make the irreversibility emerge from $M$: (i) Tomita
Takesaki's dynamic theory \cite{Takezaki 1} extended by Connes
\cite{Connes-1,Connes 2} and (ii) Kubo, Martin and Schwinger KMS
physical principle~\cite{KMS}.
\begin{itemize}
\item According to Tomita-Takesaki, if $A$ is a von Neumann algebra, there
exists a modular automorphism group $\Delta$ based on a sole parameter
$t$: $\alpha_{t}=\Delta^{-\ri t}A\Delta^{+\ri t}$ which leaves
the algebra invariant: $\rd\alpha_{t}(A)/\rd t=\lim_{\Lambda\rightarrow\infty}(2\piup \ri/h)[H_{\Lambda}A]$.
There is a canonical homomorphism from the additive group of reals
to the outer automorphism group of $A$: $B$, that is independent
of the choice of ``faithful'' state. Therefore, $\langle\Omega,B(\alpha_{t+j}A)\Omega\rangle=\langle\Omega,(\alpha_{t}A)B\Omega\rangle$,
where $(\,,)$ is the inner product.
\item The link with KMS physical constraint extends this abstract point
of view. The dynamics expression using the Kubo density matrix allows one to
change the ``shaky'' hypothesis
of thermodynamic equilibrium by giving it a dynamical expression.
KMS suggested to define the equilibrium from a correlation function $[(\gamma_{t}A)B]=[B(\gamma_{t+\ri\beta h}A)]$
allowing to associate the equilibrium with a Hamiltonian according
to $\gamma_{t}A=\exp(\ri tH/h).A.\exp(-\ri tH/h)$.
\end{itemize}
The matching of both sections leads: $\beta h=1$ which is nothing
but the emergence of a thermodynamic gauge of time while the time
variable stays perfectly reversible \cite{Connes-1}.

Starting from this analysis Jean Pierre Badiali (JPB) decided the
exploration of QM by using the local irreversible transfer joined to Feynman
\cite{Badiali 1} path integrals model based on an iconoclast existence
of 2D self-similar ``trajectories''. While this model suffers from
mathematical divergences and requires questionable renormalisation
operations, Feynman model efficiency was rapidly attested. Nevertheless,
many physicists still considered that Feynman integrals are meaningless
because the concept of trajectory should \textquotedblleft obviously\textquotedblright{}
not be relevant in QM. The discernment of JPB was to take the same
trail as Feynman, by imagining irreversible series of transition
giving birth to real self-similar paths at particles. By using a Feynman
Kac transfer formula for conditional expectation of transfer, he writes
$q(x_{0},t_{0},x,t)=\int Dx(t)\exp[-\frac{1}{\hbar}A(x_{0},t_{0};x,t)]$
in which the rules of transfer are based on a Newtonian action $A(x_{0},t_{0};x,t)=\intop\{\frac{1}{2}m[\frac{\rd x(s)}{\rd s}]^{2}+u[x(s)]\}\rd s$,
he wrote the solution required for discretizing the trajectories \cite{Badiali 2}.
These notions are not associated with any natural Hamiltonian and
require a coarse graining of the space-time. To overcome this constraint,
JPB considered the couple of functional probabilities $\phi(x,t)=\int \rd y\phi_{0}(y)q(y,t_{0};x,t)$
and $\hat{\phi}(xt)=\int q(t,x;t_{1},y)\phi_{1}(y)\rd y$ with $t_{1}>t>t_{0}$.
The evolution of a system is given by a Laplacian propagator in which
$\phi(x,t)$ is bended out by geometrical potential $u(x,t)$ according
to $\pm\frac{\partial}{\partial t}\phi(x,t)+D\Delta\phi(x,t)=\frac{1}{\hbar}u(x,t)\phi(x,t)$,
where $D=\hbar/2m$ is the quantic expression of a diffusion constant
and $\phi(x,t)$ cannot be normed. These equations are neither Chapman-Kolmogorov
equations nor Schr\"odinger like equations. Thenceforth, which physical
and geometrical meaning may we attribute to the discreet arithmetic
site on which the fractal-paths are based? How do the morphisms between
states and trajectories  determine the dynamical topos? How the
statistical or non-statistical regularizations ruling the dynamics
may smooth the experimental behavior? All these issues are open. To
solve them, JPB point of view required a new visitation of the thermodynamics
and in conformity with KMS point of view, a new definition of the
equilibrium expressed via the irreversibility of the local transfer.
To do this, he considered the class of the paths reduced to loops:
$\phi(x_{0},t_{0},x_{0},t-t_{0})$ and their fluctuations in energy.
Assuming an average energy $U$ determined by a thermostat, the overall
fluctuations are ruled by a deviation, on the one hand, from the reference
value $U$ and, on the other hand, from the number of loops concerned.
As Feynman had imagined it, an entropy function: $S_{\text{path}}=k_{\text B}\ln\int q(x_{0}t_{0},x_{0},t-t_{0})\rd x_{0}$
can be built which is ruled by the concept of path temperature $T_{\text{path}}$:
$ \frac{\hbar}{k_{\text B}}(1/T_{\text{path}})=\tau+[U-(\langle u_{K}\rangle_{\text{path}}+\langle u_{p}\rangle_{\text{path}})]\frac{\partial\tau}{\partial u}$.
The emergence of an equilibrium is figured dynamically through a critical
time scale $1/\beta h$, which possesses a strictly quantum statistical
origin merely based on loops $\tau=(\hbar/k_{\text B})T_{\text{path}}$ if it can
be assumed that the temperature of the integral of the path is none other
than the usual thermodynamic temperature. From this step, JPB finds
again the Rovelli-Connes assertions regarding thermal-time \cite{Rovelli 1}
and he proves the Boltzmann $H$-theorem. By means of subtle analysis
using the duality of the couple propagators (forward and backward
dynamics), he built a complex function $\psi,$ solution of the
Schr\"odinger equation. The thought of JPB appears as a subtle adventure
which --- inscribed in the footsteps of Richard Feynman, and implemented
from a deep knowledge of QM, thermodynamics, thermochemistry and irreversible
processes --- changes the traditional point of view and builds a perspective
that we have to analyze now, from an alternative point of view which
replaces the transport along the fractal trail by a transfer across
an interface, both perspectives being strongly related. In brief, $1/\beta$
provides a scale of energy which smooths the regime of quantum fluctuations
according to an uncertainty relation: $\Delta E=\hbar/\Delta t<1/\beta$
namely $\Delta t>\beta\hbar$. $\beta\hbar$ is the value of the
time defining the cut-off between quantum fluctuations and thermodynamic
fluctuations. The propagation function imparts a quadratic form to
the spatial fluctuations, namely $\delta x^{2}=(\beta\hbar/\partial t-1)\partial t^{2}/m$.
If $\Delta t=\beta\hbar/2$ then $\delta x^{2}=\beta\hbar^{2}/m=2\beta\hbar D\sim\delta t$,
the value that, with the reserve of taking into account the entropy constant
$k_{\text B}$, must be compared to de Broglie's length. Thus, the coarse
graining of the time will be considered as the dual of the quadratic
quantification of space, when a length in this space can be reduced
to the constraints imposed by the geometrical pattern of non-derivable
trajectories (herein with a dimension two attributed implicitly and
for quantum physical reasons to the set of Feynman paths). The aim
of this note is to show that this ``cognitive skeleton''
does not only give birth to thermal statistical time, but through
a generalization of fractal dimension, to a purely geometric irreversible
time unit: an arrow of time.

\section{Zeta function and ``$\alpha$-expon\textit{a}ntiation''}

\begin{figure}[!t]
\begin{center}
\includegraphics[scale=0.7]{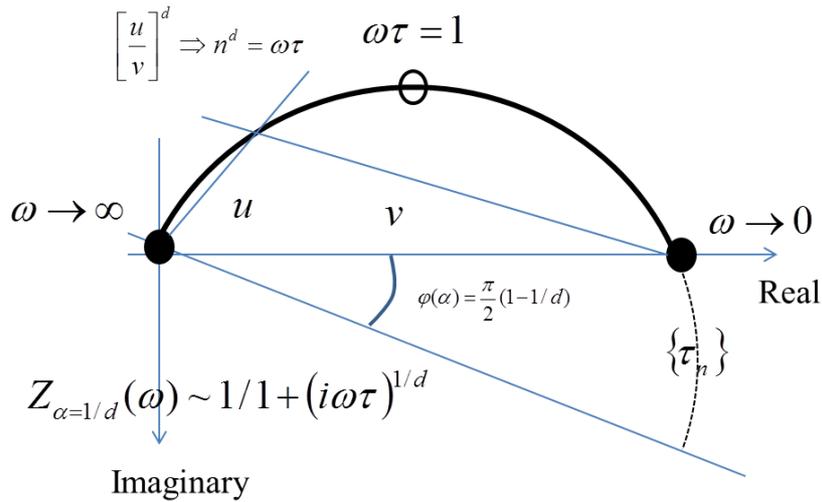}
\end{center}
\caption{(Color online) \textit{Main characteristics of exponential transfer function convoluted
by $d$-fractal geometry} $1<d\leqslant 2$. If $d$ is the rightful non integer
dimension of underlined geometry (TEISI model \cite{ALM 1982,Ttricot})
the transfer function $Z_{\alpha}(\omega)$, --- named Cole and Cole
impedance in electrodynamics \cite{ALM 1982,Penton 1,Nig 1,Jonsher 2}, ---
finds its expression in $1/d$-expon\textbf{a}ntial (with ``a''),
a kind of degenerated time function in real space shaped by a hyperbolic
geodesic close to an exponential in Fourier space. $Z_{\alpha}(\omega)$
is a hyperbolic geodesic in Fourier space. Reduced to its discrete
representation in $\mathbb{N}$ or $\mathbb{\mathbb{Q}}$ such as
$n=u/v$ or $v/u$ rational, $Z_{\alpha}^{N}(n)$ appears as the basis
for the definition of $\zeta(s)$ Riemann function ($s=\alpha+\mathit{\ri t}$)
if $\mathit{\ri t}$ expresses a mathematical fibration (figure~\ref{fig2}). The singular
points $\{0,1,\infty\}$ suggest links between $1/d$-expon\textbf{a}ntial
and Teichmuller-Grothendick absolute Galois group. Adjoined with its
categorical Kan extension $\tau$: $\{\tau_{n}\}$ obtained through the
prime-decomposition $n=\omega\tau_{n}\in\mathbb{N}$ and playing the
role of inverse Fourier transform, the above representation shapes,
--- via the pair of geodesics building the semi-circle, --- the functional
relationship between $\zeta(s)$ and $\zeta(1-\bar{s})$ \cite{ALM CMA 2010,REE,Riot 2016}. }
\label{fig1}
\end{figure}

In addition to B. Mandelbrot initial friendship, we owe to J.P. Badiali
and Professors I. Epelboin and P.G. de Gennes the first academic support
for the development of the industrial TEISI model energy transfer
on self-similar (i.e., fractal interfaces). This was at the end of seventies
shortly before the premature death of Professor Epelboin. The purpose
of this model was to explain the electrodynamic behavior of the lithium-ion
batteries which were then at the stage of their first industrial
predevelopment \cite{ALM 1982,ALM 2 1883,ALM 3 83}. The interpretation
and the patterning of electronic and ionic transfer coupled together
in 2D layered positive materials $(\text{TiS}_{2},\text{NiPS}_{3})$ are very similar
to the JPB model. The electrode is characterized by a fractal dimension
$d$ which, due to the symmetries of real space, must be such as $d\in[1,2]$.
When the fractal structure of the electrode is scanned by the transfer
dynamics through electrochemical exchanges, the electrode does not
behave like an Euclidean interface, as a straightforward separation
between two media, but like an infinite set of sheets of approximations
normed by $\eta(\omega)$ or a multi-sheet manifold, thick set of
self-similar interfaces working as paralleled interfaces \cite{Ttricot},
where $\omega$ is a Fourier variable. Each $\eta(\omega)$-interface is
tuned by a Fourier component of the electrochemical dynamics. The
overall exchange is ruled by a transfer of energy either supplied
by a battery (discharge) or stored inside the device (charge). The
impedance of positive electrode is expressed through convolution operators
coupled with the distributions of the sites of exchanges (electrode),
giving birth at macroscopic level to a class of non-integer differential
operators which take into account the laws of scaling, from quantum
scales of transfer up to the macroscopic scales of measurement. This
convolution between the discreet structure of the geometry and the
dynamics must be written in Fourier space by using an extension of
the Mandelbrot like fractal measure namely $N\eta^{d}=1$, into operator-algebra
with $N=\ri\omega\tau$ \cite{ALM 1982}. Mainly, the model emphasizes
the concept of fractal capacity (fractance) --- implicitly Choquet non-additive measure and integrals --- whose charge is ruled by the non-integer
differential equation $\ri\sim \rd^{\alpha}U/\rd t^{\alpha}$ with $\alpha=1/d$
\cite{Oldham 2,Machado}, where $U$ is the experimental potential.
In the simplest case of the first order local transfer, hence, for
canonical transfer, the Fourier transform must be expressed through
Cole and Cole type of impedance: $Z_{\alpha}(\omega)\sim1/[1+(\ri\omega\tau)^{\alpha}]$ \cite{Penton 1,Nig 1,Jonsher 2,heliod1}
which is a generalization of the exponential transfer turned by convolving
with the $d$-fractal geometry. Many other interesting expressions and
forms can be found, but being basically related to exponential operator,
the canonic form appears as seminal. The model was confirmed experimentally
in the frame of many convergent experiments concerning numerous types
of batteries and dielectric devices. JPB has advised all these developments
especially within controversies and intellectual showdowns. For instance,
even if energy storage is at the heart of all engineering purposes
\cite{ALM 2 1883,ALM 3 83}, the use of non-integer operators renders
the model accountable of the fact that energy is no longer a natural
Noetherian invariant of the new renormalizable representation. Therefore,
algebraic and topological extensions must be considered whose results
are the emergence of time-dissymmetry and of entropic-effects. Fortunately,
$Z_{\alpha}(\omega)$ clearly appears as a geodesic of a hyperbolic
space $\eta^{d}(\omega)=\frac{u}{v}$ authorizing (figure~\ref{fig1}), on
the one hand, the use of non-Euclidean metrics to establish a distance
between $\eta(\omega)$-interfacial sheets and, on the other hand, the
tricky algebraic and topological extension of the dynamics, practically
a dual fractional expression of the exponential.
\begin{figure}[!t]
\begin{center}
\includegraphics[scale=0.5]{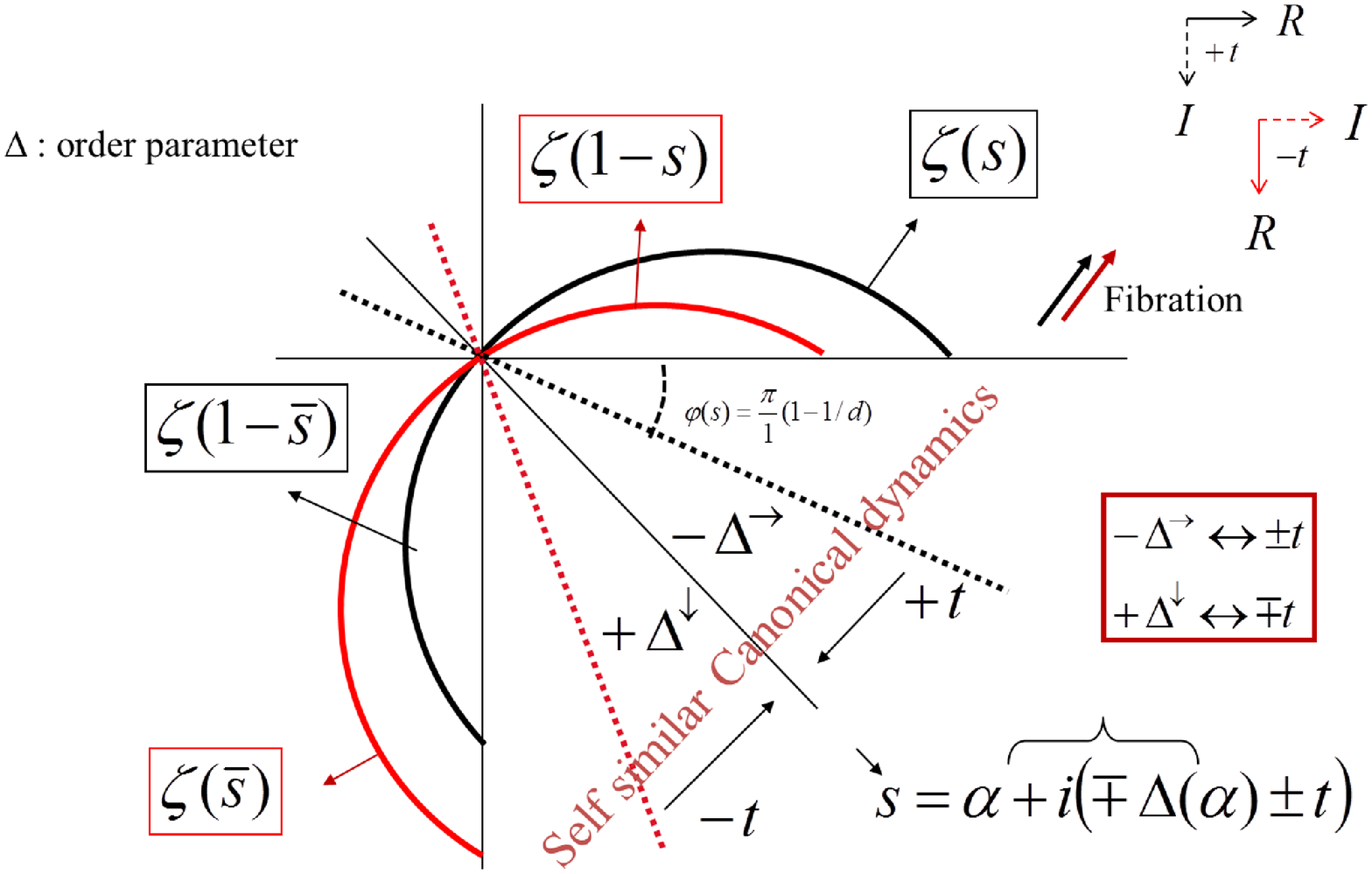}
\end{center}
\vspace{-4mm}
\caption{(Color online) \textit{Dynamical meaning of Riemann's hypothesis, phases and arrow
of time}: Riemann zeta function $\zeta(s)$ is proved to be erected
from a mathematical fibration $\pm \ri t$ based on an $\alpha$-expon\textbf{a}ntiel
represented in the oriented complex plan from $Z_{\alpha}^{N}(\omega)\cup\{\tau_{n}\}$
impedance (figure~\ref{fig1}: direct in black and inverse in grey) when $\mathbb{N}$
quantization is carried out with $s=\alpha+\ri t$. If $\alpha\protect\neq1/2$,
the constraint of phase $\pm\Delta\protect\neq0$ over the fibration
(gauge effect) is associated to the entropic properties of the dynamics
while the phase involves a dissymmetry of the fibration parameter
$\pm \ri t$, therefore, $\zeta(s)\protect\neq\zeta(\bar{s})$. If Riemann
hypothesis is validated, $\alpha=1/2$ value associated to the quadratic
self-similarity of the set of integers $\mathbb{N}\times\mathbb{N=\mathbb{N}}$,
underlined Peano geometry. This hypothesis involves $\Delta=0$
and then $\zeta(s)=\zeta(\bar{s})$. With respect to $\alpha$-expon\textbf{a}ntiel,
these properties can be expressed within a pair of vector bases: either
based on phase angle $\varphi(\alpha)=\frac{\piup}{2}(1-\alpha)$ (determinism
basis) or based on $\Delta(\alpha)$ (stochastic basis). The change
from one to the other reference must be associated to a rotation of
the dynamic referential keeping all non-linear properties $\alpha\protect\neq1/2$
of $\alpha$-geodesics. The $\Delta(\alpha)$ referential is, nevertheless,
the most fundamental of both for at least two reasons: (i) The non-commutativity
of successive $\theta$-rotations (groups) in the complex plan of
$\zeta(s)$ definition can be expressed via the equation $\theta_{1}\circ\theta_{2}=\theta_{2}\circ\theta_{1}\re^{\pm2\ri\Delta}$,
therefore, $\Delta(\alpha)=0$ implicitly provides a gauge condition
for finding again the commutativity. Due to the phase effect, the symmetry
$\zeta(s)=\zeta(1-\bar{s})$, the relation which naturally leads to $\zeta(s)=0$
as well as to the existence of Hilbert quantum mechanics states,
matching here the random distribution of primes numbers (Montgomery
hypothesis) according to the von Neumann algebra and multiplicative
self-similarity of $\mathbb{N}$: $\mathbb{N}\times\mathbb{N=\mathbb{N}}$.
The only solution to by-pass the resurgent symmetry of $t$ and to
create dissymmetry is then to introduce statistics from outside via
an external artefact: the thermostat ($t$ stays symmetric but the
random fluctuations of local process rebuild the quantum time arrow through
a so-called ``thermal time'' \cite{Badiali 2,Rovelli 1,Rovelli,arrow});
(ii) The situation is opposite if the phase moves from zeros $\Delta(\alpha)\protect\neq0$.
The loss of symmetry within the fibration keeps the internal non-commutativity
leading to an intrinsic irreversibility of the time which may be named
``arrow of time'' \cite{smolin}. Herein,
this arrow finds its origin in the open status of the geometry highlighted
through $d$ the non-integer metric of the geometry which founds the
$\alpha$-expon\textbf{a}ntiel dynamics \cite{alm98}. }
\label{fig2}
\end{figure}
The extension to
``dual $\alpha$-geodesic'' $(Z_{\alpha}^{\tau_{n}}=Z_{\alpha}\cup\{\tau_{n}\})$
shown in figure~\ref{fig1} is able to formalize the main characteristics of a
global fractional dynamics \cite{Machado} which retrieves, as we
can show below, a capability to rebuild, through the addition of entropic
factors, the contextual meaning of the physical process. We qualified
``$\alpha$-expon\textbf{a}ntiation'' (with ``a'')
this new global dynamics (figure~\ref{fig1}). This denomination integrates
the phase angle $\varphi(\alpha)$ (figure~\ref{fig1}), namely the symmetries
and inner dynamic automorphisms caused by the $d$-fractal geometry.
Let us observe that if $s=\alpha+\ri\varphi(\alpha)$ a new reference,
$0<\varphi(s)<\varphi(\alpha)$ defines a compact set $K$ able to
be considered as a base for Tomita's shift $s\rightarrow s+\ri t$ implemented
in the complex field (see below the $\mathbb{N}$ fibration). In
addition, let us observe that the expression of $\varphi(s)$ requires
a referential system which can be given either with respect to experimental
data (figure~\ref{fig1}) or with regard to an \textit{a priori} referential
obtained after a $\piup/4$ rotation, supposing the use of a Laplacian
paradigm in which $\Delta(s)$ (figure~\ref{fig2}) is used as a new expression
for phase-reference in place of $\varphi(s)$.
The reason of the relevance of this duality is an extremely deep physical meaning: according
to non-integer dynamic model if the transfer process is implemented
across a Peano curve (Feynman paths, nil co-dimension, no outer operator),
then $\alpha=1/2$ and $\Delta(s)=0,$ then the overall impedance
recovers an inverse Fourier transform and the dynamic measure fits
a probability. The traditional concept of energy recovers its practical
relevance and the space time relationship becomes coherent with the
use of the Laplacian and Dirac operators. The time used is the reversible
time of the mechanics. Conversely, if $\alpha\neq1/2$, the inverse
Fourier transform does not exist and, therefore, the traditional concept
of time vanishes, retrieving the mere arithmetic operator status implemented
in the TEISI model, namely $N=\ri\omega\tau$. Time has no longer any
straightforward usual meaning. These strange conclusions about ``time''
as well as the issues about ``energy'', left 
many academic colleagues dubious late in 1970-ies, but not JPB who
found in these issues many reasons for reviving the electrochemical
and electrodynamical concepts. Although these disturbing issues did
stay open, the TEISI experimental efficiency suggested that there
was something deeper and more fundamental behind the model; but which
thing?\,\ldots Our obstinacy to believe in the physical meaning of the
TEISI model was rewarded early this century, by discovering
that the canonical Cole and Cole impedance is closely related with
the Riemann zeta function properties \cite{ALM CMA 2010} and that
these properties are explicitly associated to the phase-locking of
fractional differential operators. $Z_{\alpha}^{N}(n)$, the integer
discretization of the Cole and Cole impedance $Z_{\alpha}(\omega)$,
is characterized by the hyperbolic-dynamic metric given by $(u/v)=1/n^{\alpha}$.
Therefore, the overall discretization of $Z_{\alpha}^{\tau_{n}}$ appears
as a possible grounding for the definition of both Riemann zeta function
$\zeta(s)$ and $\zeta(1-\bar{s})$, where $s=\alpha+\ri\varphi(\alpha)\in\mathbb{C}$,
and $1/2\leqslant \Re s<1$ \cite{ALM CMA 2010} with an emphasis given
to the arithmetic site $\{S_{\alpha}^{\tau_{n}}\}= \,_{0}^{\infty}L=\{Z_{\alpha}\cap\{\tau_{n}\}\}$
(see \cite{Machado}, page 231). An extension of the concept of time
to complex field $t\in\mathbb{R}\Rightarrow s\in\mathbb{C}$ is a
natural result of this discretization. In addition, as suggested in
recent studies \cite{Wolf,Snaith}, a heuristic reasoning about the symmetries
and automorphisms backed on $Z_{\alpha}^{\tau_{n}}$ led us to assume
(i) that the Riemann conjecture concerning the distribution of the
non-trivial zeros of zeta function $\zeta(s)=0$ could be validated
starting from physical arguing by using self-similar properties of
$\zeta(s)$ obvious from Cole and Cole impedance and recursive dynamics
\cite{ALM CMA 2010,Lapidus}, (ii) that the complex variable $s=\alpha+\ri\varphi$
also associated to the metric of the geometry through $d=1/\alpha$
accommodates, through its complex component, something of the formal
nature of the concept of arrow of time and (iii) that according to
the consequence of Montgomery hypothesis, QM states should be related
to the set of zeros, but also joined to the disappearance of above
time intrinsic arrow. We recall that the Riemann conjecture states
that the non-trivial zeros of the zeta function $\zeta(s)=0$ are
such as if $\Re s=\alpha=1/2$ (phase locking for $d=2$), namely,
in TEISI model geometrical terms, Riemann hypothesis would be related to
Peano interfaces (2D embedded geometry without any external environment).
Due to the similarity between JPB model and TEISI model which use
the irreversible transfer as test functions of the distribution of
the sites of exchanges, the morphisms concerning the scaling and the
role of the metric in this operation, leads to guess that the theory
of categories and, moreover, the theory of Topos should be hidden behind
the morphism escorted by the role of zeta function. The authors will
consider in these following paragraphs only the theory of categories.

\subsection{Universality of zeta Riemann function}

It is well known \cite{Hauet} that Riemann zeta function can be expressed by 
means of two distinct formulations: (i) additive series $\zeta(s)=\sum_{n\in\mathbb{N}}n^{-s}$
and (ii) multiplicative series $\zeta(s)=\prod_{p\in\wp}(1-p^{-s})^{-1}$.
It is also well known that any analytic function must be expressed through
additive series $f(s)=\sum_{n\in\mathbb{N}}(a_{n}s^{n})$. A duality
exists that associates $f(s)$ based on $s^{n}$ and $\zeta(s)$ based
on $n^{-s}$; under the reserve of the sign, there is an inversion
of the spaces occupied by the complex argument $s,$ and by the integer
$n$. At this step, the key point is as follows: the dual functions
can be compared using the Voronin theorem \cite{Voronin 1975,Voronin 1992,Bagchi 1982}.
This theorem states that any analytic function, for example a geodesic
on a hyperbolic manifold, can be approximated under conditions set
out, by so-called universal functions. The archetype of these functions
is precisely nothing else than Riemann zeta function $\zeta(s)$, namely:
for $K$ compact in the critical band $1/2\leqslant \alpha<1$ with a connex
complement and for $f(s)$ analytic continuous function in its interior
\textbf{without zeros on $K$}, $\forall\varepsilon>0$, $\lim\inf_{T\rightarrow\infty}(1/T)\times \text{mes}\{\tau\in[0,T];\text{max}|\zeta(s+\ri t)-f(s)|<\varepsilon\}>0$
if $t\in[0,T]$. The zeta function being used as reference, the extension
of the abscissa according to $T\rightarrow\infty$ leads to a ``crushing'' of the analytical function on the reference $\zeta(s)$. In addition,
ten years after Voronin did establish his theorem, Bagchi demonstrated
\cite{Bagchi 1982} that the validity of the Riemann hypothesis (RH)
is equivalent to the verification of the universality theorem of Voronin
in the particular case where the function $f(s)$ is replaced by $\zeta(s)$,
namely: $\forall\varepsilon>0$, $\lim\inf_{T\rightarrow\infty}(1/T)\times \text{mes}\{\tau\in[0,T];\text{max}|\zeta(s+\ri t)-\zeta(s)|<\varepsilon\}>0$.
Therefore, Bagchi's inequality asserts that the nexus of RH is the
self-similarity of Riemann function, the property explicitly content in
its link with $Z_{\alpha}^{\tau_{n}}$ \cite{ALM CMA 2010,Riot 2016,REE}.
Nevertheless, since the distribution of the
zeta zeros is unknown, it must be observed that the restriction concerning the absence of zeros
inside the compact set $K$ does not allow one to apply the Voronin
theorem to $\zeta(s)$. Therefore, if the validity of RH is related
to $\zeta(s)$ self-similarity, this property must emerge within a
theoretical status at margin with respect to the field of the analysis.
Practically, this observation urges us to consider RH as a singularity
in an enlarged field of mathematical categories and that is why the
authors suggested to introduce the category theory \cite{Cat 2 Law,Cat 3 MLa,Cat 4 Bor,Hines 1}
for handling the RH issue \cite{Riot 2016,REE}. The use of this theory
is justified for at least two reasons: (i)~according to the work of
Rota \cite{Rota}, the function $\zeta(s)$ can easily be expressed
in the framework of partially ordered sets (which forms the basis
of all standard dynamics), particular cases of categories; (ii)~since
the Leinster works \cite{Leinster 2,Cat 1 Lie}, self-similarity as
a property of a fixed point must be easily expressed by using the
language of categories. Experts in algebraic geometries will consult
with profit the reference \cite{connes 4}. The reader of this note
will be able to find in this essay and in the associated lectures \cite{Connes-1},
the reasons for which some engineers search for illustrating the profound
but also practical signification of the famous hypothesis. Both approaches
should be theoretically bonded via the existence of a renormalization
group over $Z_{\alpha}^{\tau_{n}}$ capable of compressing the scaling ambiguities
characterizing the singularities of the fractional dynamics: scaling
extension of figure~\ref{fig1} for tiling the Poincar\'{e} half plan~\cite{Machado,alm98}.

\subsection{Design of $E_{pr}$-space }

A category is a collection of objects $(a,b,\ldots)$ and of morphisms
between these objects. Morphisms are represented by arrows $(a\rightarrow b)$
which can physically account for structural analogies or dynamics
relations. Two axioms basically rule the theory: (i) an algebraic
composition of arrows $a\rightarrow b\rightarrow c$, pointing out
in the framework of set theory the homomorphism: $\text{hom}(a,c)$, and
(ii) the identity principle which accounts for an absence of any internal
dynamics of the objects $(1_{a}$: $a\rightarrow a)$. We must point out
that the compositions of ``arrows'' can
in practice be thought of as order-structures. Within the framework
of enriched categories, there is, in addition, a close link between
categories and metric-structures \cite{Cat 2 Law}. For instance,
the additive monoid $(N,+,\geqslant)$ may be substituted by $\text{hom}(a,b)$,
after the introduction of the notion of a distance through a normalization
of the length of the arrows. $\mathbb{N}$ is naturally associated
to the additive law (\textit{construction}) which provides, through
the monoid $(N,+)$, an ordered list of its elements $[1,2,3,\ldots, n,n+1,\ldots]$,
but it is also associated through the monoid $(N,\times)$ to a distinct
order structure. The question of matching both monoids makes it possible to consider it as a mere arithmetic  issue, but the order associated
to $(N,\times)$ or $(N,\div)$ must over here, --- the\textit{ partition}
of the set $\mathbb{N}$, --- be defined in the following way: $p<q$
if and only if $p$ \textit{divides} $q$. \textit{The order is, therefore,
only partial} because, as it is well known, any integer may be written
in a unique way according to $n=\prod_{i}p_{i}^{r_{i}}$ with $p_{i}$
prime and $r_{i}$ integer while $i$ scans a finite collection of
$n\in\mathbb{N}$; the order appearing through the set of $p_{i}$
is mainly different from the order of the set of $(N,+)$. By taking
the logarithm, one obtains  $\log(n)=\sum_{i}r_{i}\times \log p_{i}$
for all $n$. Mathematically, $\mathbb{N}$ with partial order is a
``\textit{lattice}'': any pair of elements $x$ and $y$ has a
single smallest upper bound, in this case, the LCM (Lowest
Common Multiple) and a single GLD (Greatest Lower Divisor).
The total order structure itself also constitutes a lattice
for which the operators max and min can be substituted
by LCM and the GLD. The elements of a lattice can be quantified by
associating a value $v(p)$ for each $p$ in such a way that for $p\geqslant q$
we have $v(p)\geqslant v(q)$. According to Aczel theorem \cite{Aczel 1,aczel 2,Asso 1},
there is always a function possessing an inverse such that the linear
ordered discrete set of values makes it possible to match the partial
order associated with the multiplicative monoid $(N,\times)$ and
the order associated to $(N,+,\leqslant )$. In the above particular case, this
result takes the following form: both monoids $(N,+,\leqslant )$ and $(N,\times)$
are in correspondence by means of a logarithm, so that: $\log[\text{LCM}(p,q)]=\log p+\log q-\log[\text{GLD}(p,q)]$.
The dissymmetry between \textit{construction} and \textit{partition}
explains the role of non-linear logarithm function, hence, the
paradigm of exponential function in the physics of close additive
systems, while, conversely, the treatment of non-extensive systems
stays always an open issue. Although very elementary, these characteristics
of  $\mathbb{N}$ invite us to introduce a space of countable
infinite dimension $E_{pr}$ which is characterized by an
orthogonal vectorial basis indexed by the quantities $\log p_{i}$,
where  $p_{i}$ is any prime integer and wherein the vectors
have a finite number of integer coordinates $r_{i}$, the other coordinates
being reduced to zero \cite{Riot 2016,REE}. Indeed,
$E_{pr}$ is remarkably well adapted to a linearization of the self-similar
properties expressed from the discrete framework of $\mathbb{N}$.
The orthogonal character of the basic axis accounts for the fact that
the set of prime numbers constitutes an anti-chain upon the partially
ordered set associated with the divisibility, hence, $\mathbb{Q}$
the set of rational numbers, such as defined above. The space $E_{pr}$
corresponds to the positive quadrant of a Hilbert space in which the
norm of unity vector is equal to $\log p_{i}$. It is then easy to
introduce the scaling factor using a parameter based on the complex number
$-s\in\mathbb{C}$. At coordinate point, $r_{i}$ is then associated with
the coordinate $-s\times r_{i}$. The space obtained by applying the
scaling function $s$ may be noted as $N(s)$ \cite{REE}. The construction
of this kind of space using the logarithmic function is all the more
relevant in that its inverse, i.e., the exponential function, can be applied
in return. Therefore, the total measure of the exponential operator
can then be easily computed  upon the set of integer points constrained
by a complex power law $n^{-s}\in N(s)$ on $E_{pr}$ for any chosen
parameter $s$. This operation gives birth to zeta Riemann function
$\zeta(s)=\sum_{n\in\mathbb{N}}n^{-s}$ which finds, therefore, in $E_{pr}$
its natural mathematical sphere of definition. The zeta Riemann function
is the total measure of the exponentiation operator applied upon the
set $N(s)$ when expressed in $E_{pr}$, and $\zeta(s)$ is, therefore,
merely the trace of this operator in $E_{pr}$: $\zeta(s)=\Tr_{E_{pr}}\{\exp[-s\log N(s)]\}$.

At this step, it is interesting to confront the above analysis to
quantum mechanics. For example, one can observe that Montgomery-Odlyzko
hypothesis (MOH) could be based on a specific interpretation of $E_{pr}$
space. Let us remind that Montgomery considers the identity of distributions
between the zeros pair correlations of the Riemann zeta function and
the eigenvalue peer correlations of the Hermitian random matrices
\cite{Snaith}. The conjecture asserts the possibility of regularizing
divergent integrals by using a Laplace operator whose spectra are based
upon the $\mathbb{N}$ ordered series of vectors $0\leqslant \lambda_{1}\leqslant \ldots\leqslant \lambda_{n}\leqslant \ldots<\lambda_{\infty}$.
Then, one can define the zeta spectral function according to the equation
$\zeta_{\lambda}(s)=\sum_{n\in N}(1/\lambda_{n})^{s}$. This function
is only convergent for $s\in\mathbb{R}$ but it has an extension in
the complex plane. For the hermitian operator $H,$ we have $\zeta_{\lambda}(s)=\Tr[\exp(-s\log H)]$
while $\det H=\prod_{n\in N}\lambda_{n}$. Therefore, with respect
to $\zeta(s)$, the description requires an introduction of the concept
of energy according to $\log(\det H)=\Tr(\log H)$. Then, the Mellin transform
of the kernel of ``\textit{heat equation}'' can then be expressed
using: $\zeta(s)=\int_{0}^{\infty}t^{s-1}\Tr[\exp(-tH)]\rd t$ with $\Tr[\exp(-tH)]=\sum_{n\in N}\exp(-t\lambda_{n})$
leading to the Riemann hypothesis. But, being upstream of this specific
problem, by highlighting the role of any partial order even for $\alpha\neq1/2$,
$E_{pr}$ founds and, within a certain meaning, generalizes the implicit
assumptions in MOH. $E_{pr}$ overcomes the limiting role
of Laplacian operator and the role Hermitian hypothesis which
implicitly and \textit{a priori} imply the additive properties of the systems
concerned, or in other words, admit \textit{a priori} the validity
of the Riemann hypothesis [the existence of well defined random states
associated to the zeros of zeta function: $\Re s=\alpha=1/2$]. The
categorical link, described above for any values of $s$, between
$(N(s),+,\leqslant)$ and $(N(s),\div)$ [i.e., $(N(s),\times)$] referred,
respectively, to the total order (forward construction) and to the partial
order (backward partition) is well adapted for dealing with non-additive
systems, namely a dynamical conception of being a steady state
of arithmetic exchanges, without any additional hypothesis regarding $\Re s$.
Obviously, the above conception can be narrowed to additive systems or steady
state if $\alpha=1/2$. According to this overall point of view,
the categorical matching between \textit{construction} and \textit{partition}
which gives birth to a renormalization group, might be physically expressed
through gauge constraints, namely, intrinsic automorphisms required
for closing the system over itself \cite{Connes 2,connes 4}. Many
other essential properties of multi-scaled systems could be unveiled
by formalizing the theory from $E_{pr}(s)$ space, even if very singular
interesting properties arise when, according to Riemann hypothesis,
$\Re s=1/2$, one introduces additional specific symmetries in $E_{pr}$
such as $\zeta(s)=\zeta(1-\bar{s})$. In general, whatever the $\alpha$
value, the function $\zeta(s)$ is the total measure of the exponentiation
operator on the support space $N(s)$ at a certain scale $s$ while
$\zeta(s)$ is constrained by Bagchi inequality based upon a time-shift
$s$ to $s+\ri t$, very identical to the one used in Tomita and KMS relations.
In order to analyse a possible analogy between both approaches, it
appears then necessary to analyze how the space $N(s)$ behaves under
the shift when $\ri^{2}=-1$.

\subsection{$E_{pr}$ fibration}

Let us consider the parameter $s=\alpha+\ri\varphi$ variable in a compact
domain $K\in\mathbb{C}$ such that, $\alpha\in[1/2,1]$ and $0\leqslant \varphi\leqslant \varphi_{\text{max}}(\alpha)\leqslant \piup/4$
(figure~\ref{fig2}). According to Borel-Lebesgue theorem, a compact domain
in $\mathbb{C}$ is a closed and bounded set for the usual topology
of $\mathbb{C}$, directly inherited from the topology associated
with~$\mathbb{R}$. The $K$ bounded character is essential for backing
the reasoning based on the shift from $s$ to $s+\ri t$. Indeed, by
choosing a parameter $t\in\mathbb{\mathbb{N}}$ as a value sufficiently
high with respect to the diameter of the domain $K$, the shift from
$s$ to $s+\ri t$ makes it possible to create a translation of the domain $K$
\cite{REE} with a creation of copies of $K$: $K_{t}$ capable of avoiding
any overlapping if a relevant period $t=\tau$ is rightly chosen.
Thus, $t$-shifts uplift a fiber above $K$. In the frame of $\alpha$-expon\textbf{a}ntial
representation, this characteristic may be practically applied for
folding the dynamic and zeta function if, for instance, $K$ is associated
to the field of definition of $Z_{\alpha}^{\tau_{n}}$, while $Z_{\alpha}(\omega)$
is used to root $\zeta(s)$ on the set $\{\alpha,\varphi_{\text{max}}(\alpha)\}$.
Let us observe in advance that $\re^{\pm \ri t}$ implements the fibration
by starting from the gauge-phase angle $\varphi_{\text{max}}$ (figure~\ref{fig2}).
This way for understanding the fibration is equivalent to replacing
the additive operation ($s$ to $s+\ri t$) by a Cartesian product.
If we now replace $s$ with $s+\ri m\tau$, where $m$ scans the countless infinite
set of integers; the reciprocal image along the base change is then
the fiber product of space $N(s)$ by a discrete straight line defined
by $\ri\times m\times\tau$. The total space is characterized by $N(s)\times\{\ri\times m\times\tau\}\simeq N(s)\times N(s)\simeq N(s)$.
Thus, the change of the basis does not realize anything else but the bijection
$\mathbb{\mathbb{N}\times\mathbb{N}=N}$, characterizing a well-known
quadratic self-similarity characteristic of the set of integers. The
self-similar characteristics of $\mathbb{N}$ can be approached by
using a particular class of polycyclic semigroups or monoids \cite{fibre ,Nivat,Howies,Lawson 1}.
They are representable as bounded linear operators of a traditional
Hilbert space,  of type $N(s)$ herein. The change of base consists in
introducing such a semigroup realizing a fibration based on the self-similarity
$N(s)\times N(s)=N(s)$, or a partition within subspaces with co-dimension
1. Each sheet corresponds to the space above the variable $s+\ri\times m\times\tau_{0}$. The value of the Riemann function $\zeta(s+\ri\times m\times\tau)$
is obtained as the total measure of the exponential operator on each
sheet, namely this value is a truncation of $\zeta(s)$. This truncation
is the basis of Riemann hypothesis. Let us observe that $m$ which
is obtained after a rearrangement of the numerical featuring corresponding
to the isomorphism $\mathbb{\mathbb{N}\times\mathbb{N}=N}$ imposes
a distribution of points $\alpha+\ri\times m\times\tau$ that, in the
complex plan, does not mesh the total order given from $(N,+,\leqslant )$.
Above each complex number $s\in K$ and along the fiber, an appropriate
category exists in $E_{pr}$ based upon both initial and terminal
object $N(s)$ \cite{SmyPlot,Belaiche1,Lambek} leading to the folding
of the $\mathbb{N}$ and, therefore, to the second different order. The
``disharmony'' between both orders involved by the relation $\mathbb{N}{}^{2}=\mathbb{N}$
has its equivalent in the TEISI model when the previous self-similarity
is expressed by $\eta^{2}\times(\ri\omega\tau)=1$. The interface of
transfer is then a Peano interface, where the complex variable $\ri$
expresses the fibration and $\omega\tau=n\in N(s)$ is used for the
computation of $\zeta(s)$. However, via the operator general equation
$\eta^{d}\times(\ri\omega\tau)=1$, this \textquotedblleft disharmony\textquotedblright{}
is notified by tagging the sign of $t$ through a phase factor $\ri^{1/d}$
generally different from $\ri^{1/(1-d)}$. Therefore, one should distinguish
at least two cases:

\textit{First case: time symmetry and the absence of junction phase}.
To avoid dissymmetry of the phase at boundary, the singularity of the phase
angle must be canceled, namely, $\varphi(\alpha)=\piup/4$, $\Delta=0$
(figure~\ref{fig2}). The dynamics basis must be expressed through $Z_{1/2}^{\tau_{n}}(s)$,
which gives birth to a folding of $\zeta(1/2\pm \ri t)$. RH becomes
associated to the expression of the invariance of $t$ under a change
of sign. In terms of phase transition, $t$ is a parameter of order
and $\alpha=1/2$ is the tag which points out a singularity of ``order'' within the ``disorder'' ruled by $\alpha\neq1/2$,
$\zeta(s)\neq0$. The main order which must be considered whatever
$s\in\mathbb{C}$ is naturally given through $\zeta(s)=0$.

\textit{Second case: existence junction phase.} If we take into account
the fact that TEISI relation is more general than the quadratic one
and must be considered under its general form: $\eta^{d}(\ri\omega\tau)=1$
namely $\eta\times\ri^{\alpha}(\omega\tau)^{\alpha}=1$, $\ri^{\alpha}$
introduces a critical phase angle when fibration is implemented $\mathbb{\mathbb{N}\times_{\varphi}\!\mathbb{N}=N}$.
If $t$ is the physical time parameter, this relation proves the
existence of an arrow of time emerging from the underlying fractal
geometry, if the metric of this geometry requires an environment.
The main mathematical issue revealed by the controversies is then
our capability or not of reducing the fractal dynamics to a stochastic
process, namely $(\omega\tau)^{\alpha}\rightarrow(\omega'\tau')^{1/2}$.
Provided we take into account the phase angle, the presence
of $\Delta(s)$ suggests that this transformation could be rightful
if a thermodynamical free energy were considered (Legendre transform).
The question which must be also addressed within an universe characterized
by an $\alpha$-exponantiation with $\alpha\neq1/2$, namely, the
disappearance of perfectly defined Hilbert states, concerns the class
of groupoids capable of replacing Hilbert-Poincar\'{e} principle. These troublesome
issues occupied the latest scientific conversations I had with Jean
Pierre Badiali.

\section{\textit{Pro tempore} conclusion regarding an arrow of time }

The definition of a concept of time requires a unit which, within a
progressively restrained point of view from $\mathbb{R}$ to $\mathbb{N}$
(or $\mathbb{Q}$) should match the set $[0,1]\cup]1,\infty]$ onto $[0,1]$,
namely a basic loop. Backed on the TEISI model and a general $\alpha$-geodesic
which provides a dynamic hyperbolic meaning to Riemann zeta functions,
the use of $E_{pr}$ space and $\mathbb{N}$ self-similar category,
offers the chance to understand the ambivalence of the concept of
physical time. The ambivalence, that unfolds through a complex value of
time, may be expressed using a pair of clocks HL and HG. HL
is related to the additive monoids. Indeed, as shown from $\alpha$-expon\textbf{a}ntial
model, $n\in N(s=\alpha+\ri\varphi)$ can be associated with the scanning
of a hyperbolic distance $l_{\text{HL}}=(u/v)^{d}=(1/n)^{d}$ defined on
the geodesic $Z_{\alpha}(\omega)$ according to the additive monoid
$(N,+,\leqslant )$ (figure~\ref{fig1}). The computed hyperbolic ``path integral''
\cite{ALM CMA 2010} is no other than $\zeta(s)=\sum_{n\in\mathbb{N}}n^{-s}$.
Therefore, the evolution of the $n$ along the geodesic is ruled by
pulsing $n$. This reversible time can be easily extended to $\mathbb{Q}$
and to $\mathbb{R}$ as usually done. However, by arguing the concept
of time from such a dynamic context, one reveals the existence of a
second tempo on HG: $t=\omega\tau$ with $t\in\mathbb{N}$ capable
of being tuned to the first clock only in the frame of von Neumann (operator)
algebra, taking into account an appropriate phase chord $u/v=(\ri\omega\tau)^{1/d}$.
Indeed, through the fractal metric, determination of the absolute
values of this tempo and, therefore, the matching of both approaches
implies the critical role of the phase $\varphi(\alpha)$ [or $\Delta(\alpha)$
if referred to $1/2$-geodesic], the phase which has an impact, without any
possible avoidance, on the sign of the fibration $\mathbb{\mathbb{N}\times_{\varphi}\!\mathbb{N}=N}$.
The second clock HG can be tuned in upon the pulse of the first
one HL, like $\mathbb{N}\times\mathbb{N}$ must be tuned on $\mathbb{N}$
through a product, by adjusting the edges. Practically, two situations
must be taken into consideration:
\begin{itemize}
\item $\alpha=1/2$, in the frame of the dynamic model, the base of the fibration
is the $1/2$-expon\textbf{a}ntial geodesic which is a degenerate
form of the dynamics characterized by the removal of any exteriority.
This form is associated to Riemann hypothesis. The Laplace transform
of non-integer operator exists. The energy fills in its usual Noetherian
meaning. The spectrum of the operator applied on $\mathbb{N}$ can
be built upon the set of prime numbers giving birth to the category
of Hilbert eigenstates. Due to the quadratic form, the chord of both
clocks can be easily obtained. The characteristics of Laplacian natural
equation may account for this tuning which originates in the quadratic
self-similar structure of $\mathbb{N}$: $\mathbb{N}^{2}$$=\mathbb{N}$
also expressed in the TEISI equation $\ri\omega\tau.[\eta(\omega\tau)]^{2}\simeq1$.
The irreversibility of the time can only have an external origin;
the thermal time unit is then nothing else than the unit of time associated
to the Gaussian spatial correlations meshed by the temperature associated
to an external thermostat, which, by locking the type of fluctuations,
smooths the Peano interfacial geometry via a stochastic process. Fortunately,
for energy efficiency, the engineering of batteries is not based on this
principle.
\item $\alpha\in]1/2,1],$ the dynamics is based on the incomplete $\alpha$-expon\textbf{a}ntial
geodesic. There is not any natural Laplace transform for such geodesics
and the spectrum over $\mathbb{N}$ cannot provide any simple basis
for the representation of inner automorphisms joined together in a
``bundle'' $\{\tau_{n}\}$ which assures
a completion, but an entanglement when the closing of the degrees of freedom
becomes the heart of the physical issues. Fortunately, an integral
involution can be built whose minimal expression can be based upon
the hybrid complex set of couples $\left\{\zeta(s)\oplus\zeta(1-\bar{s});\zeta(\bar{s})\oplus\zeta(1-s)\right\}$
in which $\oplus$ expresses the disjoint sum of the basic ``geodesics''.
$\{\zeta(1-\bar{s});\zeta(1-s)\}$ plays the role usually devoted
to the inverse Laplace transform. These couples of functions that
take into account, through $s$ and $\bar{s}$, the sign of the fibration
(rotation in the complex plane of geodesics) assure the tuning of
the complex dynamics and fix the status of the time taking into
account the sign of fibration. This analytical context brings the two main issues to light.

\begin{itemize}
\item The question of commensurability of the couple clocks HL and HG
which, as above, can be physically tuned by using a thermal regularization
(entropy production). This regularization can be based upon a Legendre
transform defined from the upper limit of the $\alpha$-geodesics.
This transform is allowed by the possible thermodynamic involution
between $\alpha$-geodesics and $1/2$-geodesics whose equation $|\Delta(\alpha)|+|\varphi(\alpha)|=\piup/4$
provides the insurance. This involution might explain the dissipative
auto-organizations, well known in physics as well as the existence
of some optimal values of fractal dimensions in irreversible processes,
especially the critical dimensions, $d_{a}=4/3$ and $d_{g}=7/4$.
\item Infinitely more meaningful is the presence of the phase angle $\pm\Delta(\alpha)\neq0$
which imposes an absolute distingue between both possible signs of
the parameter of fibration and a non-commutativity of the associated
operators for folding. In this context, and exclusively in this context,
the reversibility of the cyclic operators, along the fiber must be
expressed by $t_{1}.t_{2}=t_{2}.t_{1}\re^{\pm2\ri\Delta}$, non-commutative
expression from which the notion of ``arrow of time''
takes on an irrefutable geometrical signification and, herein, an interfacial
physical meaning. The irreversibility is then clearly based on the
freedom of a boundary phase, namely the initial conditions, when $\mathbb{N}(\times\mathbb{N})_{\varphi(\alpha)}$
the fibration realized the matching between \textit{construction}
and \textit{partition}. Intrinsic irreversibility should then originate
from the boundary property. It is then in the thermodynamic framework that
the $\pm$, namely, the difference between ``future''
and ``past'', must be analyzed, by
assigning the emergence of time-energy to the distingue between the
work and the heat. The arrow of time justifies the practical emergence
of the distingue of HL and HG while Legendre transformation
can ensure the mathematical validity of the passage from one to the
other of the notions.
\end{itemize}
\end{itemize}
These last elements very exactly summarize the content of the ultimate
discussions shared with my friend Jean Pierre Badiali. Starting from
Feynman analysis, his talent had assumed that the reversibility of
the time usually required for representing the dynamics of quantum
processes should be a very specific case (closed path integrals) of
a more general situation (local dissipation, open path integrals and
non-extensive set) fundamentally based on the local irreversibility and
ultimately complicated by the convolution with a set of non-differential
discrete paths. The problem of the ``open loops''
and their non-additive properties, will stay as an open issue for
him. He assured with courage this uncomfortable position during his
last ten years of research, exploring with me all trails capable of conferring
a coherence to his mechanical approach. His vision matched, at least
partially, the main-stream choices of quantum mechanics according to
which the basis of macroscopic irreversibility should be the result
of a statistical scaling closure, settled by the contact with a thermostat
or an experimenter. In this paradigmatic framework, the concept of
thermal time has no other physical origins than these externalities.
As we have tried to show synthetically in this note, our last exchanges
concerned the possibility of passing this option for building a
hybrid point of view using the role of zeta functions. He attempted
without success to introduce this function in his own model but he
understood the deep signification of Riemann hypothesis to describe
complex systems which possess well defined internal states. We had
imagined our writing a book together, titled ``Issues
of Time''. The disappearance of Jean Pierre has not only
suspended this project, but has left us scientifically fatherless
in front of (i) the complexity of all physical open questions, (ii)
the urgency of assuring science that should never reduce to the only
technosciences and, furthermore, (iii) that the research of all truths
still hidden within a shadow preserves for ever its human dynamics.

\section*{Acknowledgements}

The authors would like to thank Materials Design Inc \& SARL (Dr.
E. Wimmer), the Federal University of Kazan (Prof. Dr. D. Tayurskii
the Professor Abe (MIE University Japan) and the office of the Universit\'{e}
du Qu\'{e}bec \`{a} Chicoutimi, Institut Franco-Qu\'{e}b\'{e}cois in Paris (Dr. S.
Raynal) for the support of these studies. Gratefulness for the ISMANS
Team, especially Laurent Nivanen, Aziz El Kaabouchi, Alexandre Wang
and Fran\c{c}ois Tsobnang for 16 years of fundamental research (1994--2010)
about non-extensive systems and quantum simulation, in collaboration
with Jean Pierre Badiali as member of Scientific committee.

\ukrainianpart
\title{Від стріли часу в квантовому підході Бадіалі до динамічного значення гіпотези Рімана}
\author{П. Ріот\refaddr{label1}, A. лє Меоте\refaddr{label1,label2,label3} }
\addresses{
\addr{label1}
Франко-Квебекський інститут, Париж, Франція
\addr{label2}
Відділи фізики та інформаційних систем, Казанський федеральний університет,
\\
Казань, Татарстан, Російська Федерація
\addr{label3}
Проектування матеріалів, Монруж, Франція
}

\makeukrtitle

\begin{abstract}
Новизна останніх наукових робіт Жана-П'єра Бадіалі  бере початок з квантового підходу, який базується на (і) поверненні 
до поняття траєкторій
 (траєкторії Фейнмана), а також на (іі)  необоротності квантових переходів. Ці іконокластичні варіанти знову встановлюють гільбертіан
і алгебраїчну точку зору фон Неймана, маючи справу зі  статистикою за циклами.  Цей підхід надає зовнішню термодинамічну першопричину поняттю квантової одиниці часу (термальний час Ровеллі Коннеса). Це поняття, базис для квантування, виникає тут 
як   простий критерій розрізнення між квантовим режимом і термодинамічним режимом. Мета цієї статті є розкрити зміст останніх п'яти 
років наукових дискусій, націлених з'єднати в  когерентну схему
як уподобання і роботи Жана-П’єра, так і роботи авторів цієї статті на основі   гіперболічної геодезії, і об'єднуючу роль дзета-функції Рімана.
Хоча ці варіанти не представляють жодних протиріч, тим не менше, вони породжують власну стрілу часу, інакшу ніж термальний час. 
Питання фізичного змісту гіпотези Рімана як основи квантової механіки, що було в центрі наших останніх дискусій, є суттю цієї статті.

\keywords інтеграли за траєкторіями, диференціальні рівняння в часткових похідних, дзета-функції, стріла часу
\end{abstract}

\end{document}